\documentclass[sigconf, noacm, dvipsnames]{acmart}

\usepackage{url}
\usepackage{soul}
\usepackage{xcolor}

\newcommand{\projectName}[1]{AIM}
\newcommand{\projectFullName}[1]{Alignment-in-Memory}

\setcopyright{none}
\renewcommand\footnotetextcopyrightpermission[1]{}

\begin{document}

\title{A Framework for High-throughput Sequence Alignment using Real Processing-in-Memory Systems}

\author{
    Safaa Diab\,$^1$ \quad
    Amir Nassereldine\,$^1$ \quad
    Mohammed Alser\,$^2$ \quad
    Juan G\'{o}mez Luna\,$^2$ \\
    Onur Mutlu\,$^2$ \quad
    Izzat El Hajj\,$^1$
}

\affiliation{
\institution{
    \vspace{10pt}
    $^1$American University of Beirut \quad
    $^2$ETH Zürich
}
}

\begin{abstract}

Sequence alignment is a memory bound computation whose performance in modern systems is limited by the memory bandwidth bottleneck.
Processing-in-memory architectures alleviate this bottleneck by providing the memory with computing competencies.
We propose \projectFullName{} (\projectName{}), a framework for high-throughput sequence alignment using processing-in-memory, and evaluate it on UPMEM, the first publicly-available general-purpose programmable processing-in-memory system.

Our evaluation shows that a real processing-in-memory system can substantially outperform server-grade multi-threaded CPU systems running at full-scale when performing sequence alignment for a variety of algorithms, read lengths, and edit distance thresholds.
We hope that our findings inspire more work on creating and accelerating bioinformatics algorithms for such real processing-in-memory systems.

Our code is available at \href{https://github.com/safaad/aim}{https://github.com/safaad/aim}.

\end{abstract}

\maketitle
\pagestyle{plain}

\section{Introduction}\label{sec:INTRODUCTION}

One of the most fundamental computational steps in most genomic analyses is \emph{sequence alignment}.
This step is formulated as an \emph{approximate string matching} (ASM) problem~\citep{navarro2001guided}, which typically uses a dynamic programming (DP) algorithm to optimally calculate the type, location, and number of differences in one of the two given genomic sequences.
Such sequence alignment information is typically needed for DNA sequence alignment, gene expression analysis, taxonomy profiling of a multi-species metagenomic sample, rapid surveillance of disease outbreaks, and many other important genomic applications.

DP-based alignment algorithms, such as Needleman-Wunsch (NW)~\citep{needleman1970general} and Smith-Waterman-Gotoh (SWG)~\citep{gotoh1982improved}, are computationally-expensive as they have quadratic time and space complexity (i.e., O($n^2$) for a sequence length of $n$). 
It is mathematically proven that subquadratic alignment algorithm cannot exist~\citep{backurs2015edit}. 
Recent attempts for improving sequence alignment tend to follow one of three key directions:
    (1) accelerating the DP algorithms using hardware accelerators,
    (2) accelerating the DP algorithms using heuristics and limited functionality, and
    (3) reducing the workload for alignment by filtering out highly dissimilar sequences using pre-alignment filtering algorithms.
Comprehensive surveys have been done on these existing attempts~\citep{alser2020accelerating, alser2020technology, alser2022going}.

The first direction accelerates \emph{exact} sequence alignment using existing hardware devices, such as SIMD-capable multi-core CPUs, GPUs, and FPGAs, or using to-be-manufactured devices, such as application-specific integrated circuits (ASICs).
One of the most recent alignment algorithms, the wavefront algorithm (WFA)~\citep{marco2020fast}, indeed benefits from acceleration via SIMD~\citep{marco2020fast, marco2022optimal}, GPUs~\citep{aguado2022wfa}, and FPGAs~\citep{haghi2021fpga}.
Parasail~\citep{daily2016parasail}, BWA-MEM2~\citep{vasimuddin2019efficient}, and mm2-fast~\citep{kalikar2022accelerating} all exploit SIMD-capable and multi-core CPUs to accelerate sequence alignment in read mapping.
SillaX~\citep{fujiki2018genax} provides an order of magnitude of acceleration through specialization, but requires fabricating their architecture designs into real hardware, which is costly and performed at a semiconductor fabrication facility.

The second direction includes limiting the functionality of sequence alignment to performing only edit distance calculation, as in Edlib~\citep{vsovsic2017edlib}, or limiting the number of calculated entries in the DP table, as in windowing/tiling the DP table~\citep{turakhia2018darwin, rizk2010gassst, arlazarov1970economical} and the X-drop algorithm~\citep{zhang2000greedy} implemented in one of the versions of KSW2~\citep{li2018minimap2}.
This second direction is not mutually exclusive from the first and can also benefit from acceleration via SIMD~\citep{li2018minimap2}, GPUs~\citep{ahmed2020gpu}, and FPGAs~\citep{banerjee2018asap}.

The third direction is to \emph{early} and \emph{quickly} detect any two dissimilar genomic sequences, which differ by more than a user-defined edit distance threshold, and exclude them from being aligned as their alignment result is not useful.
Pre-alignment filtering usually saves a significant amount of time by avoiding DP-based alignment \emph{without} sacrificing alignment accuracy or limiting the algorithm functionality as demonstrated when using even basic CPU implementations~\citep{rasmussen2006efficient, xin2013accelerating, rizk2010gassst, alser2020sneakysnake}.
Pre-alignment filtering can also benefit from hardware acceleration~\citep{xin2015shifted, alser2017gatekeeper, alser2017magnet1, kim2018grim, alser2019shouji, alser2020sneakysnake, cali2020genasm}.

Regardless of which of these three directions are followed to improve performance, sequence alignment remains a fundamentally memory-bounded computation with a low data reuse ratio~\citep{gupta2019rapid,10.1145/3383669.3398279,cali2020genasm,lavenier2020variant}.
Sequence alignment implementations suffer from wasted execution cycles due to the memory bandwidth bottleneck faced when moving data between the memory units and the computing units (e.g., CPUs, FPGAs, GPUs).
This bottleneck exists because of the large disparity in performance between the compute units and memory units in modern computing systems.
Following Moore's law, the number of transistors in processors has been doubling about every two years, leading to an exponential increase in the power of the processor cores~\citep{moore1998cramming}.
However, memory performance did not scale comparably which has made the cost of transferring data between the main memory and the CPU more expensive than the computations to be performed by a CPU instruction on the data~\citep{mutlu2019processing, mutlu2020modern}.

Processing-in-memory (PIM) architectures alleviate the data movement bottleneck of modern computing systems by providing the memory with computing competencies~\citep{mutlu2019processing, mutlu2020modern,wen2017rebooting, hajinazar2021simdram, hajinazar2021simdram1, ferreira2022pluto, mansouri2022genstore,puma,panther,huang2021mixed}.
PIM has been used to improve the performance of a wide variety of memory-bound computations, including sequence alignment as is the case in RAPID~\citep{gupta2019rapid}, BioSEAL~\citep{10.1145/3383669.3398279}, GenASM~\citep{cali2020genasm}, and SeGraM~\citep{cali2022segram}.
However, these PIM-based sequence alignment solutions rely on emerging technologies that require either major changes to existing hardware or fabricating new hardware chips that are specially designed for the subject algorithm.
These limitations pose a critical barrier to the adoption of PIM in sequence alignment.

UPMEM is the \emph{first} publicly-available programmable PIM system in the market~\citep{devaux2019true}.
The UPMEM architecture integrates conventional DRAM arrays and general-purpose cores called DPUs into the same chip.
This architecture allows computations to be performed near where the data resides, which reduces the latency imposed by data movement.
UPMEM systems have been used to accelerate memory-bound applications such as database index search, compression/decompression, image reconstruction, genomics, and many others~\citep{DBLP:journals/corr/abs-2105-03814,gomez2021benchmarking,gomez2022benchmarking,giannoula2022sparsep,lavenier2016mapping,lavenier2016blast,lavenier2016dna,diab2022high,lavenier2020variant,nider2021case,zois2018massively}.

Our goal is to evaluate the suitability of real PIM systems for accelerating sequence alignment algorithms.
To this end, we introduce \projectFullName{} (\projectName{}), a framework for PIM-based sequence alignment that targets the UPMEM system.
\projectName{} dispatches a large number of sequence pairs across different memory modules and aligns each pair using compute cores within the memory module where the pair resides.
\projectName{} supports multiple alignment algorithms including NW, SWG, GenASM, WFA, and WFA-adaptive.
Each algorithm has alternate implementations that manage the UPMEM memory hierarchy differently and are suitable for different read lengths.

We evaluate \projectName{} on a real UPMEM system and compare the throughput it can achieve with that achieved by server-grade multi-threaded CPU systems running at full scale.
Our evaluation shows that a real PIM system can substantially outperform CPU systems for a wide variety of algorithms, read lengths, and edit distance thresholds.
For example, for WFA-adaptive, the state-of-the-art sequence alignment algorithm, \projectName{} achieves a speedup of up to 2.56$\times$ when the data transfer time between the CPU and DPUs is included, and up to 28.14$\times$ when that data transfer time is not included.
These speedups to sequence alignment can translate into substantial performance improvements for widely used bioinformatics tools such as minimap2, where alignment can consume up to 76\% of the time~\citep{kalikar2022accelerating}, or BWA-MEM, where alignment can consume up to 47.2\% of the time~\citep{vasimuddin2019efficient}.
Our results demonstrate that emerging real PIM systems are promising platforms for accelerating sequence alignment.
We hope that our findings inspire more work on creating and accelerating bioinformatics algorithms for real PIM systems.

\section{System and Methods}\label{sec:appr}

In this section, we provide an overview of PIM and the UPMEM PIM architecture, (Section~\ref{sec:pim}).
We then describe the overall workflow of our PIM-based sequence alignment framework (Section~\ref{sec:workflow}), how we manage data within the UPMEM PIM memory hierarchy (Section~\ref{sec:wramvsmram}), and how each of the different sequence alignment algorithms are supported within our framework (Section~\ref{sec:alignment-algorithms}).

\subsection{PIM and the UPMEM PIM Architecture}\label{sec:pim}

\begin{figure}[t]
    \centering
    \includegraphics[width=\columnwidth]{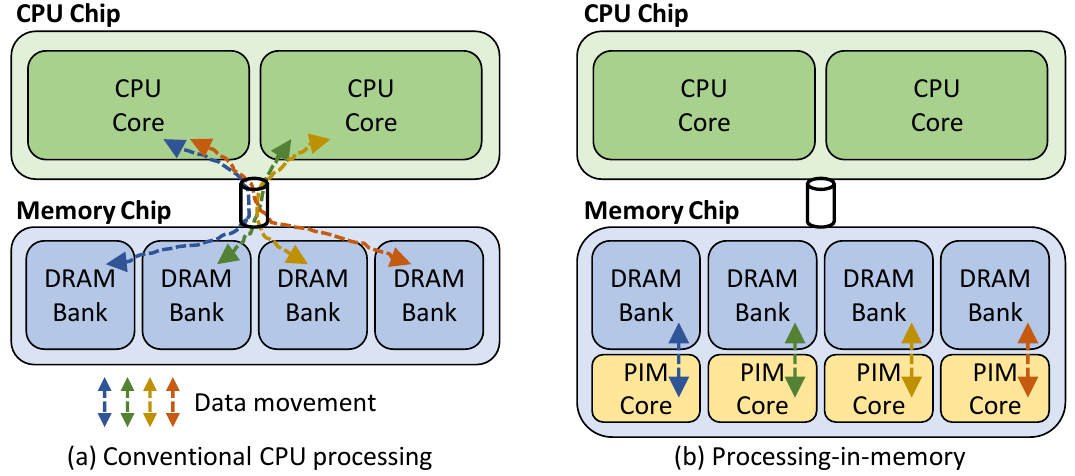}
    \caption{Comparison of conventional CPU processing and processing-in-memory (PIM)}\label{fig:pim-illustration}
    \vspace{-10pt}
\end{figure}

Fig.~\ref{fig:pim-illustration} compares the organization of conventional CPU processing systems to PIM systems.
In conventional systems, illustrated in Fig.~\ref{fig:pim-illustration}(a), data resides in dynamic random access memory (DRAM) which is typically organized into multiple DRAM banks.
This data is transferred to the CPU cores where computations are performed on the data.
If the same data is reused for many computations, the cost of moving that data from memory to the CPU cores is amortized.
However, if the data is not reused, the cost of moving the data is much higher than the cost of the computations on that data.
In this case, the CPU cores will be idle most of the time waiting for memory accesses to complete, and the movement of data between the CPU cores and the memory becomes a major performance bottleneck.

In a PIM system, illustrated in Fig.~\ref{fig:pim-illustration}(b), small PIM cores are placed in the memory chip near the memory banks.
These PIM cores are typically much less powerful than CPU cores at performing computations, however they are much faster at accessing data from memory because of their proximity.
Hence, if only a few computations are performed on the data, it is cheaper to perform these computations in the PIM cores than to move the data all the way to the CPU cores.
In this case, programmers would load their code onto the PIM cores and execute it there where the data can be fetched quickly, instead of executing it on the CPU cores.
Sequence alignment algorithms are well-suited for such a system because there are usually a few computations performed for each data element accessed from the intermediate data structures (e.g.,~entries of the DP table in NW).

The UPMEM PIM architecture~\citep{devaux2019true} is the first publicly-available general-purpose programmable (\url{https://sdk.upmem.com}) processing-in-DRAM architecture.
An UPMEM system consists of a set of UPMEM DIMM modules plugged alongside main memory (standard DDR4 DIMMs) and acting as parallel co-processors to the host CPU.
An UPMEM module is a DDR4-2400 DIMM with 16 PIM-enabled chips, where each chip consists of eight general-purpose processing PIM cores called \emph{DRAM Processing Units} (\emph{DPUs}).
Each DPU is coupled with a 64MB DRAM bank called \textit{main RAM} (\textit{MRAM}).
A current UPMEM system supports up to 20 UPMEM modules, which is equivalent to 2560 DPUs and 160GB of memory.
A DPU is a 32-bit RISC processor with a proprietary Instruction Set Architecture (ISA), which can potentially run at 500 Mhz.
Each DPU has 24 hardware threads that share a 24KB instruction memory (IRAM) and a 64KB scratchpad memory which is also called a working RAM (WRAM).
The threads also share the 64MB MRAM bank coupled with the DPU.
DPUs cannot communicate with other DPUs or access data outside their own MRAM bank.
The host CPU transfers data between the main memory and the MRAM banks, and coordinates communication and synchronization across DPUs if needed.

\subsection{Overall Workflow}\label{sec:workflow}

\begin{figure}[t]
    \centering
    \includegraphics[width=\columnwidth]{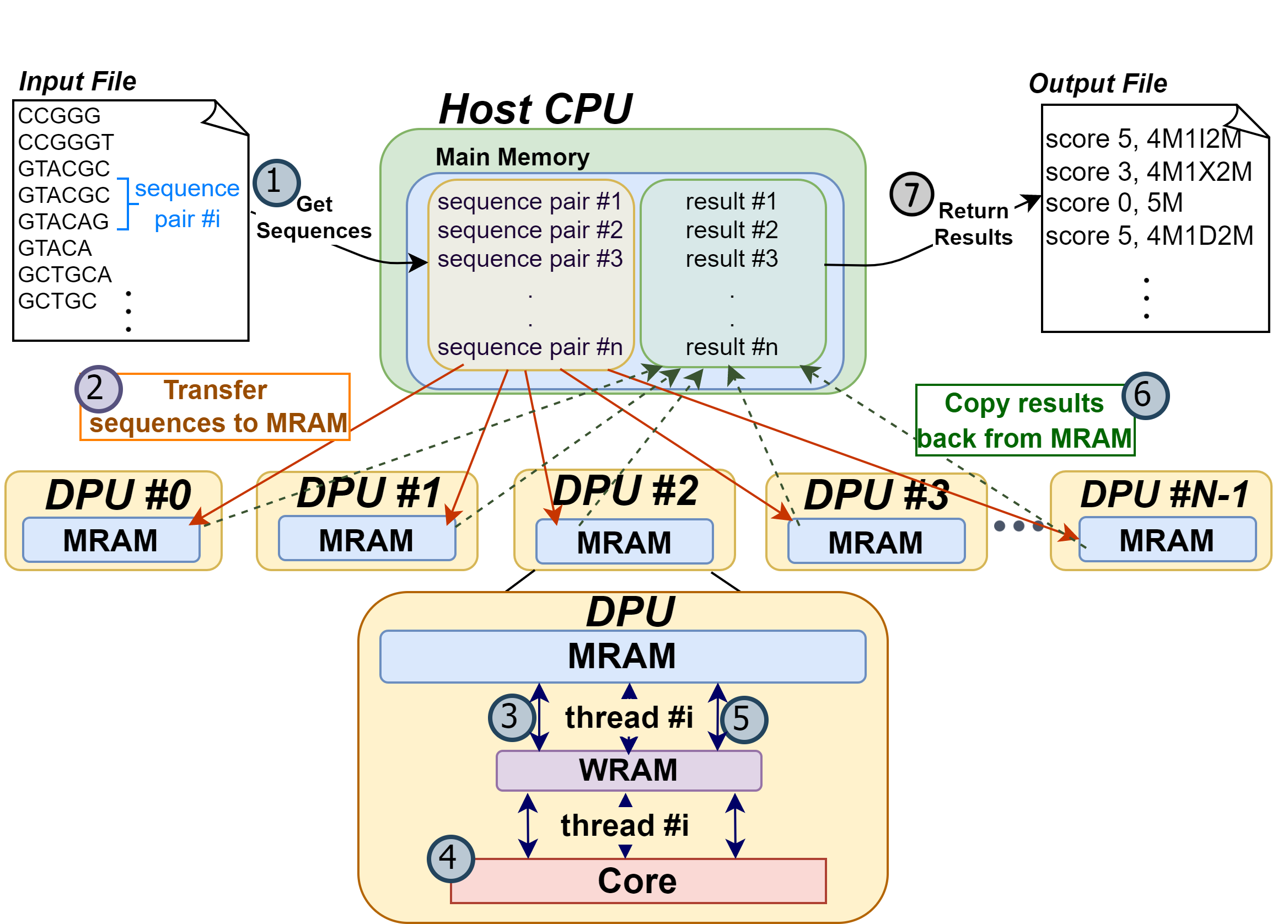}
    \caption{Overall workflow of \projectFullName{} (\projectName{})}
    \label{fig:workflow}
\end{figure}

Fig.~\ref{fig:workflow} illustrates our framework's overall workflow.
In Step (1), we read the sequence pairs from an input file on disk and store them to the main memory.
In Step (2), we transfer the sequence pairs from the main memory to the UPMEM DIMMs, distributing them evenly across the MRAM banks of the different DPUs.
We use parallel transfers so that the sequence pairs are written to multiple MRAM banks simultaneously, thereby optimizing the transfer latency.

Next, we launch the DPU kernels and have each thread in each DPU work independently to align a set of sequence pairs.
This parallelization scheme avoids inter-DPU and inter-thread synchronization, which can be expensive in the UPMEM system~\citep{DBLP:journals/corr/abs-2105-03814}.
In Step~(3), each DPU thread performs a DMA transfer to fetch a sequence pair from MRAM and store it in WRAM.
In Step (4), the DPU thread aligns the sequence pair and extracts the alignment operations using traceback.
Our framework supports five different alignment algorithms: NW, SWG, GenASM, WFA, and WFA-adaptive.
In Step (5), the DPU thread performs a DMA transfer to write the alignment score and operations to MRAM.
The thread then moves on to process the next sequence pair, repeating Steps (3), (4), and (5) until all sequence pairs have been processed.

Finally, once all DPUs finish execution, we retrieve and save the alignment results.
In Step (6), we transfer the alignment score and operations of each sequence pair from the UPMEM DIMMs to main memory using parallel transfers.
In Step (7), we load the results from main memory and write them to an output file on disk.

\subsection{Data Management}\label{sec:wramvsmram}

One important aspect of implementing alignment algorithms on the UPMEM PIM architecture efficiently is data management.
Recall from Section~\ref{sec:pim} that an UPMEM DPU has access to two memory spaces for data: a 64KB WRAM and a 64MB MRAM.
WRAM is fast and is accessed via loads and stores, whereas MRAM is slow and is accessed via DMA transfers to and from WRAM.
Our framework always places the full sequence pair to be aligned and the full alignment result in WRAM for fast access because these data items are small.
However, the intermediate data structures used by the alignment algorithms are relatively large.
For this reason, it may not be possible to fit the entire intermediate data structure for each DPU thread in WRAM while supporting a large enough number of DPU threads to efficiently utilize the DPU pipeline.
In this case, the constrained WRAM capacity can act as a limiting factor to parallelism.

\begin{figure}[b]
    \includegraphics[width=\columnwidth]{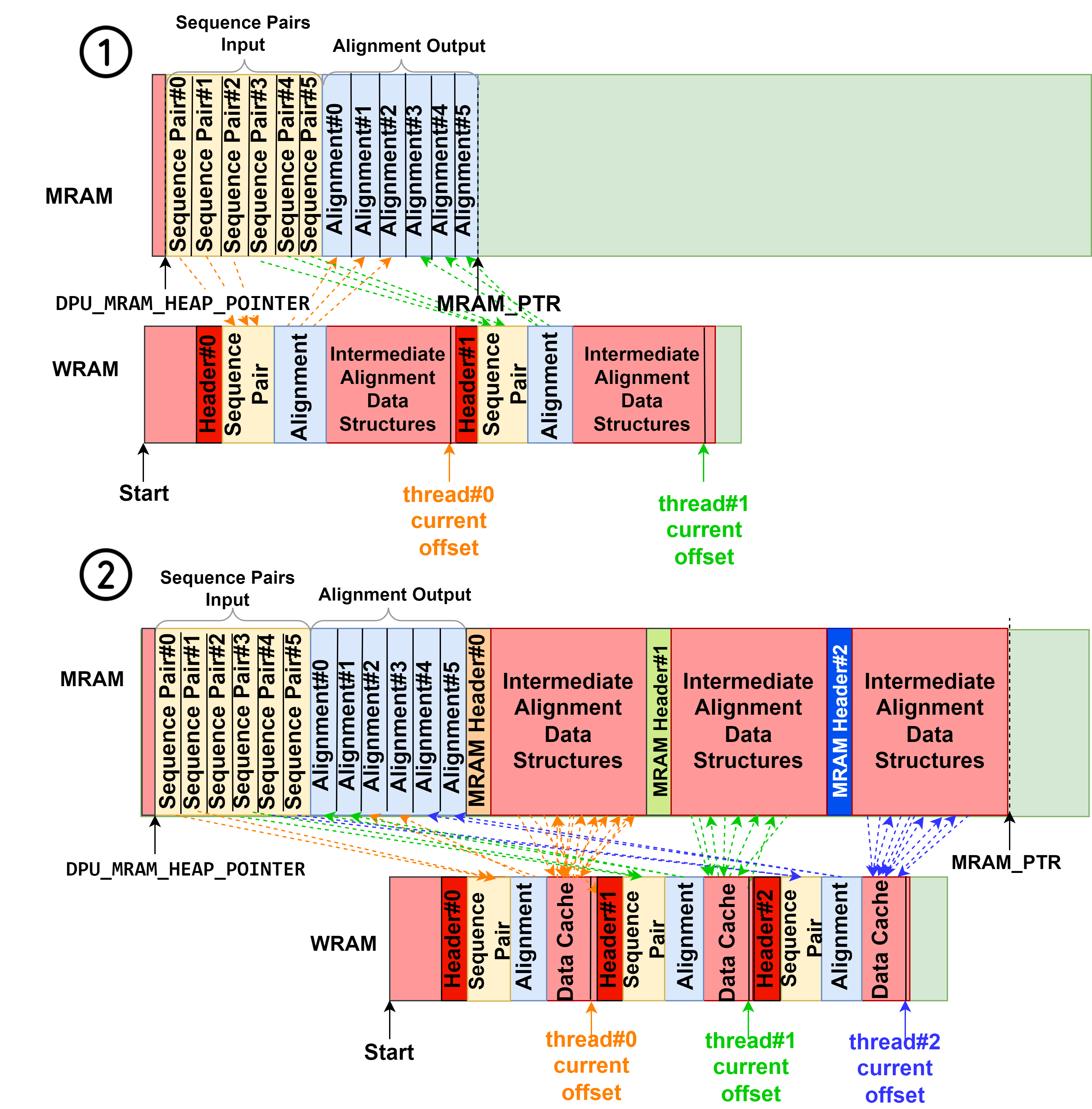}
    \caption{Example of using WRAM only (top) or using both WRAM and MRAM (bottom) for intermediate alignment data structures}
    \label{fig:mram-illust}
\end{figure}

To tackle this trade-off, we provide two alternative implementations of each alignment algorithm.
This first implementation, illustrated in Fig.~\ref{fig:mram-illust} (top), places the entire intermediate data structure for each DPU thread in WRAM, thereby prioritizing fast load/store access to the intermediate data structures.
However, as the aligned sequences get larger, the WRAM capacity begins to constrain the number of DPU threads that can be launched, which causes the DPU pipeline to be underutilized.
The second implementation, illustrated in Fig.~\ref{fig:mram-illust} (bottom) places the intermediate data structure for each DPU thread in MRAM and transfers the parts of the data structure that need to be accessed to WRAM on-demand.
With the WRAM capacity less of an issue, this approach enables using a larger number of DPU threads to better utilize the DPU pipeline.
However, it incurs longer access latency to the intermediate data structures.
The user of our framework can easily select which implementation they would like to use, and our framework automatically adapts the number of DPU threads launched depending on the implementation selected as well as the choice of alignment algorithm, read length, and error rate.
We evaluate the trade-off between these two implementations in Section~\ref{sec:results-wram-vs-mram}.

Another important aspect of data management is performing dynamic memory allocation efficiently.
NW, SWG, and GenASM have fixed-size data structures that are allocated in WRAM or MRAM up-front.
However, WFA and WFA-adaptive rely on dynamic memory allocation which is performed frequently as the algorithms run.
Dynamic memory allocation is needed for allocating the wavefront components, which vary in size at run time depending on the read length and similarity.
The UPMEM SDK has an incremental dynamic memory allocator which allocates memory incrementally from beginning to end and then frees it all at once.
However, it is not suitable for our purpose because it is shared by all threads so it requires synchronization across threads to perform allocation, which degrades performance.
To overcome this issue, we provide our own per-thread custom memory allocator to perform low-overhead dynamic memory allocation in the WFA and WFA-adaptive algorithms.
The allocator also ensures that allocations are properly aligned such that they can be used in DMA transfers between WRAM and MRAM.

\subsection{Supported Alignment Algorithms}\label{sec:alignment-algorithms}

\subsubsection{Needleman-Wunsch (NW)}

NW~\citep{needleman1970general} computes the alignment of sequences using a DP-table.
In the implementation that only uses WRAM for intermediate alignment data structures, we have each DPU thread allocate its entire DP-table in WRAM, fill it, perform the traceback, and send the alignment result to MRAM.
The size of the DP-table is $m \cdot n$ where $m$ and $n$ are the lengths of the two aligned sequences.
The data type of the DP-cells is $int16$.
Therefore, the WRAM memory consumption per thread consists of the DP-table ($m \cdot n \cdot sizeof(int16)$), the sequence pair, and the traceback operations.
In this case, the 64KB WRAM capacity limits the maximum read length to 175bp, where one DPU thread is executed consuming 61KB of WRAM.
On the other hand, in the implementation that uses MRAM for intermediate alignment data structures, we store the DP-table of each DPU thread in MRAM and use WRAM to store the neighboring DP-table cells of the current cell being computed.
In this case, the 64MB MRAM capacity limits the maximum read length to around 4Kbp.
NW uses the linear gap model to compute the alignment score.
In our evaluation, we set the scoring parameters to $a=0$ (match cost), $x=3$ (mismatch cost), and $e=4$ (deletion/insertion cost).

\subsubsection{Smith-Waterman-Gotoh (SWG)}

SWG~\citep{gotoh1982improved} resembles NW, but uses an affine gap model which treats opening new gaps and extending existing gaps differently.
To do so, SWG uses three DP-tables -- Matching (M), Insertion (I), and Deletion (D) -- each of size $m \cdot n$.
In the implementation that only uses WRAM for intermediate alignment data structures, we have each DPU thread store all three DP-tables in WRAM.
Since the memory usage of SWG is higher than that of NW, the maximum read length that can be used in this case is only 100bp.
On the other hand, in the implementation that uses MRAM for intermediate alignment data structures, we store the three DP-tables of each DPU thread in MRAM and use WRAM to store the neighboring DP-table cells of the current cell being computed.
In this case, the maximum read length that can be used is around 2.5Kbp.
We physically store the three tables as a single table where each cell has three consecutive values.
By doing so, we can transfer cells from all three tables with the same DMA transfer, thereby amortizing the cost of transferring data from MRAM over fewer DMA transfers.
In our evaluation, we set the scoring parameters to $a=0$ (match cost), $x=3$ (mismatch cost), $o=4$ (deletion/insertion opening cost), and $e=1$ (deletion/insertion extension cost).

\subsubsection{GenASM}

GenASM~\citep{cali2020genasm} is a recently proposed alignment algorithm that modifies and adds a traceback method to the bitap algorithm~\citep{10.1145/135239.135243,wu1992fast}.
It uses the affine gap model and takes as an input the maximum number of edit distances ($k$) allowed while computing the alignment.
In the implementation that only uses WRAM for intermediate alignment data structures, we have each DPU thread use WRAM to store the pattern bit-mask for each character in the alphabet, two status bit-vectors to hold the partial alignment between subsequences of the sequences in the pair, and four intermediate bit-vectors for each edit case (matching, substitution, deletion, and insertion).
On the other hand, in the implementation that uses MRAM for intermediate alignment data structures, we store the pattern bit-mask and the two status bit-vectors in MRAM and transfer parts of them to WRAM as needed.
However, the four intermediate bit vectors are still allocated in WRAM since they are small in size.
In our evaluation, we set $k$ according to the read length and error rate used. 

\subsubsection{Wavefront Algorithm (WFA)}

WFA~\citep{marco2020fast} is the state-of-the-art affine gap alignment algorithm that computes exact pairwise alignments efficiently using wavefronts in the DP-table.
Each wavefront represents an alignment score, and the algorithm finds successive wavefronts (i.e., computes increasing-score partial alignments) until reaching the optimal alignment.
Hence, the complexity of WFA is $O(n \cdot s)$, where $s$ is the alignment score.
As the alignment score $s$ increases, WFA takes longer to execute and consumes more memory because it spans more diagonals.
For this reason, WFA-adaptive~\citep{marco2020fast}, a heuristic variant of WFA, reduces the number of the spanned diagonals by eliminating outer diagonals that are unlikely to lead to an optimal alignment.
Note that both WFA and WFA-adaptive have linear space complexity, which makes them representative of other linear space sequence alignment algorithms~\citep{myers1988optimal,durbin1998biological} that could also benefit from our framework.

In our framework, we provide implementations of both WFA and WFA-adaptive.
In the implementations that only use WRAM for intermediate alignment data structures, we have each DPU thread use WRAM to store all the wavefront components.
On the other hand, in the implementations that use MRAM for intermediate alignment data structures, we store all the wavefront components in MRAM, and keep the addresses of the components in WRAM so they can be found when needed.
To compute a new wavefront component $WF_s$, a DPU thread transfers from MRAM to WRAM only the components it needs.
After computing $WF_s$, the DPU thread transfers the result from WRAM to MRAM.
In our evaluation, we set the scoring parameters to $a=0$, $x=3$, $o=4$, and $e=1$.

\section{Evaluation}\label{sec:EVAL}

\subsection{Experimental Setup}\label{sec:setup}

We compare the performance of our proposed framework to a mutli-threaded CPU baseline that uses OpenMP to align multiple sequence pairs in parallel.
The baseline CPU implementations of NW, SWG, WFA, and WFA-adaptive are taken from the original WFA repository~\citep{marco2020fast}, and the baseline CPU implementation of GenASM is taken from the GenASM repository~\citep{cali2020genasm}.

We evaluate our PIM implementations on an UPMEM system with 2560 DPUs (20 UPMEM-DIMMs) running at 425MHz.
We evaluate the CPU implementations on three different \textit{server-grade} CPU systems shown in Table~\ref{tab:cpus}.
We use execution time as the basis for comparison, which includes the wall clock time for performing data transfer, alignment, backtrace, and CIGAR string generation.

\begin{table}[h]
    \centering
    \resizebox{\columnwidth}{!}{
        
\begin{tabular}{cccc}
    \hline
    \textbf{System} & 1 & 2 & 3 \\
    \hline
    \textbf{CPU} & Intel Xeon Silver 4215 & Intel Xeon Gold 5120 & Intel Xeon E5-2697 v2 \\
    \textbf{Process node} & 14 nm & 14 nm & 22 nm \\
    \textbf{Sockets} & 2 & 2 & 2 \\
    \textbf{Cores} & 16 & 28 & 24 \\
    \textbf{Threads} & 32 & 56 & 48 \\
    \textbf{Frequency} & 2.50 GHz & 2.20 GHz & 2.70 GHz \\
    \textbf{L3 cache} & 22 MB & 38 MB & 60 MB \\
    \textbf{Memory} & 256 GB & 64 GB & 32 GB \\
    \textbf{CPU TDP} & 170 W & 210 W & 260 W \\
    \hline
\end{tabular}

    }
    \caption{CPU systems used in the evaluation}
    \label{tab:cpus}
    \vspace{-20pt}
\end{table}

We use real and synthetic datasets to evaluate our proposed framework, as shown in Table~\ref{table:datasets}.
The real datasets are short read-reference pairs generated using minimap2~\citep{10.1093/bioinformatics/bty191} by mapping the datasets (\url{https://www.ebi.ac.uk/ena/browser/view}) mentioned in Table~\ref{table:datasets} to the human reference genome GRCh37~\citep{church2011modernizing}.
The synthetic datasets are long sequence pairs simulated using the synthetic data generator provided in the WFA repository~\citep{marco2020fast}.

\begin{table}[ht]
    \centering
    \resizebox{\columnwidth}{!}{
        
\begin{tabular}{lll}
\hline
Read Lengths & Edit Distances & Description\\
\hline
 100 & 0-5\% & Real, Accession\# ERR240727 \\
 150 & 0-5\% & Real, Accession\# SRR826460 \\
 250 & 0-5\% & Real, Accession\# SRR826471 \\
 500, 1000, 5000, 10000 & 0-5\% & Synthetic~\citep{marco2020fast}\\
\hline
\end{tabular}

    }
    \caption{Datasets used for the evaluation}
    \label{table:datasets}
    \vspace{-20pt}
\end{table}

\subsection{CPU Performance}\label{sec:results-cpu}

\begin{figure}[h]
    \centering
    \includegraphics[width=\columnwidth]{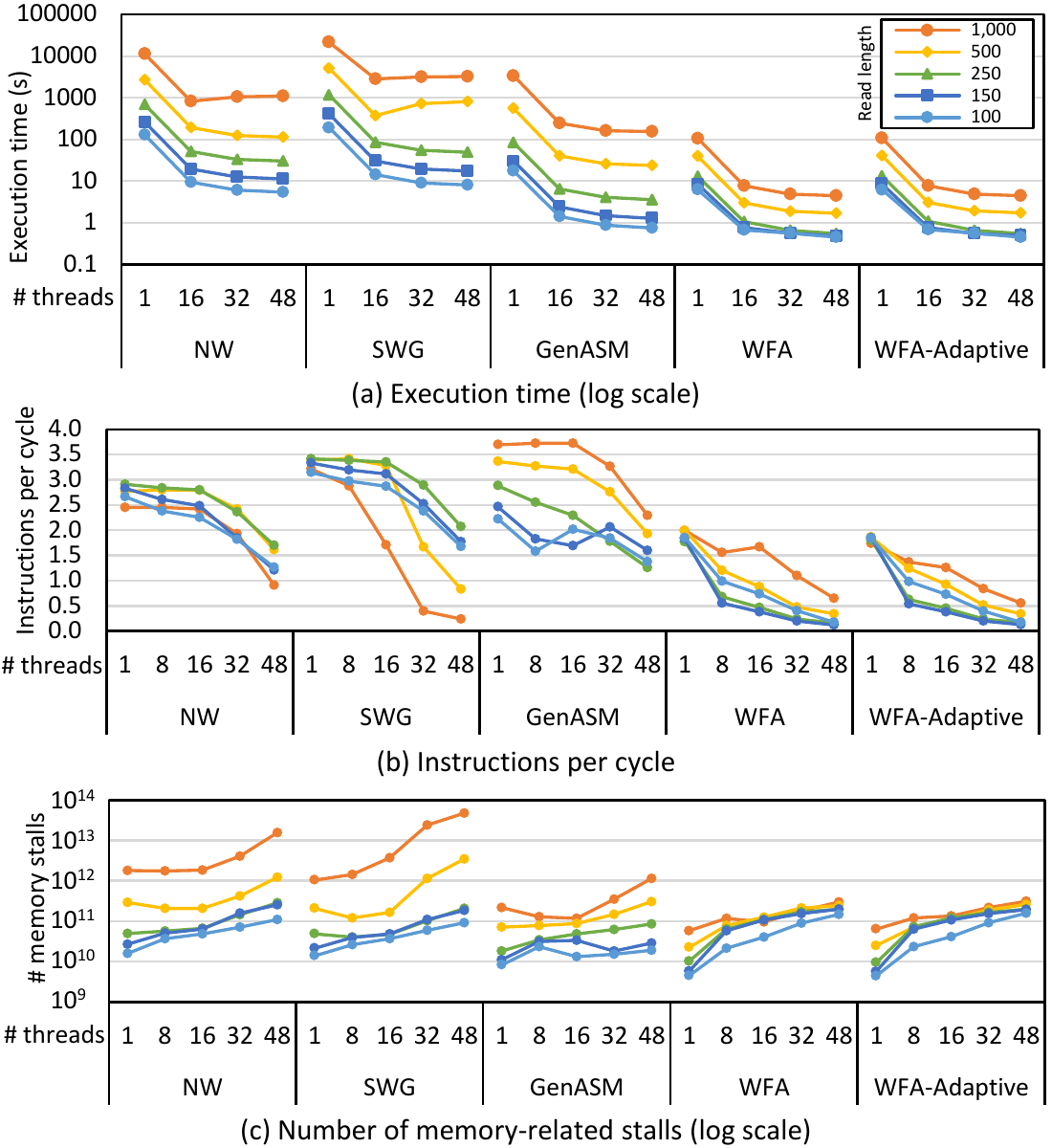}
    \caption{Evaluation of CPU implementations on the best performing CPU system (Intel Xeon E5) while aligning five million sequence pairs }\label{fig:results-cpu}
\end{figure}

Fig.~\ref{fig:results-cpu}(a) shows how the execution times of the CPU implementations scale with the number of CPU threads while aligning five million sequence pairs using different alignment algorithms and read lengths, and an edit distance of 1\%.
The results are reported for the Xeon E5 CPU system, which is the best performing CPU system as we show in Section~\ref{sec:results-pim}.
Our key observation is that the CPU implementations face limited performance improvement as the number of CPU threads grows.
To investigate the cause of this limited performance improvement, we profile the application using \texttt{perf}.
Fig.~\ref{fig:results-cpu}(b) shows that as the number of CPU threads increases, the instructions executed per cycle by each thread decreases.
Hence, the CPU is increasingly underutilized when more threads are added.
To further understand what is causing the CPU to be underutilized, Fig.~\ref{fig:results-cpu}(c) shows that as the number of CPU threads increases, the cycles spent by the CPU stalling while waiting for memory requests increases.
Hence, the limited performance improvement is caused by the inability of the memory to serve memory requests quickly enough.
These results demonstrate the importance of using PIM to overcome the memory bandwidth bottleneck faced by sequence alignment applications.

\subsection{PIM Performance versus CPU Performance}\label{sec:results-pim}

\begin{figure*}[ht]
    \centering
    \includegraphics[width=\textwidth]{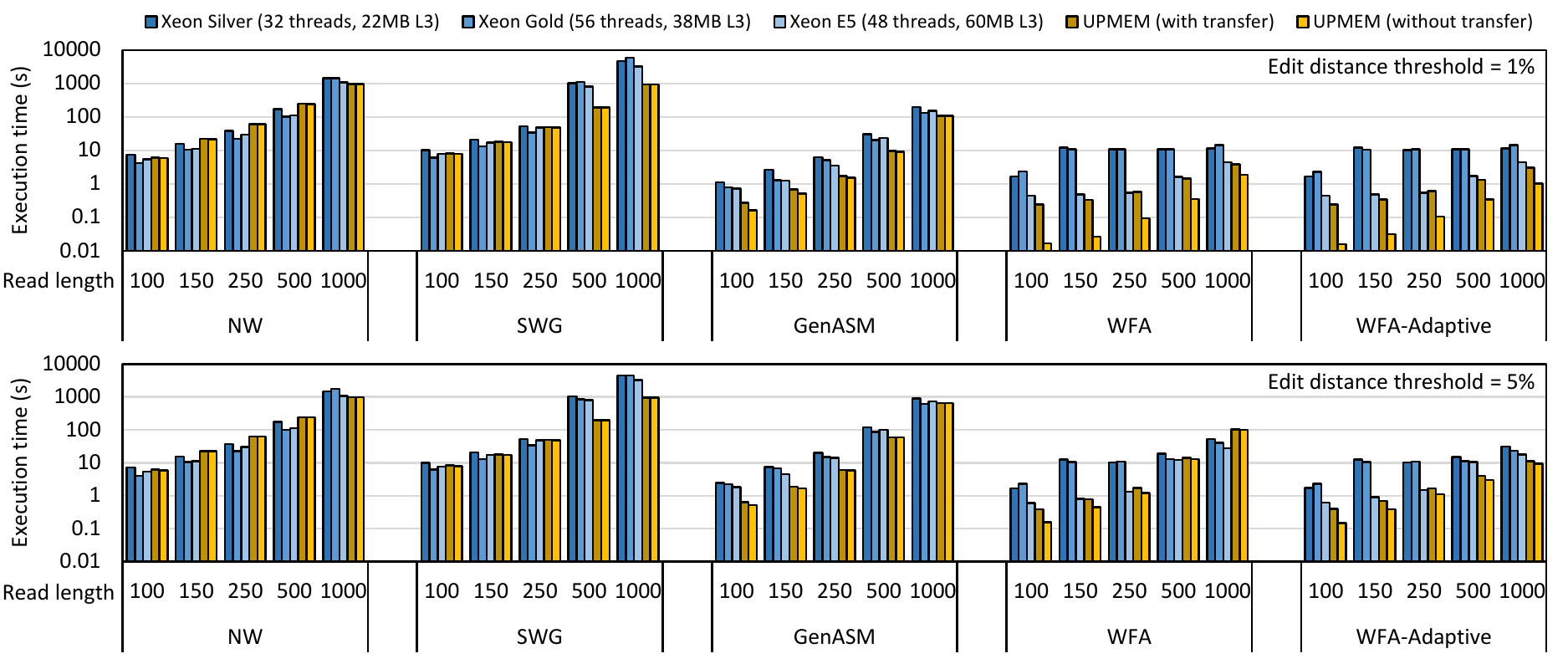}
    \caption{Execution time of our framework while aligning five million sequence pairs using different alignment algorithms, read lengths, and edit distances compared with three CPU baselines}\label{fig:results-performance}
\end{figure*}

Fig.~\ref{fig:results-performance} shows the execution time of our framework aligning five million sequence pairs using different alignment algorithms, read lengths, and edit distances.
We report the execution time both with and without the data transfer time between main memory and the UPMEM DIMMs.
Based on these results, we make four key observations.

The first observation is that among the three CPU baselines, the best performing baseline in the majority of cases is the Xeon E5 CPU system.
Despite not having the largest number of threads, this baseline has the largest L3 cache.
This result demonstrates the memory-boundedness of the sequence alignment problem, where having a larger L3 cache is favored over more having more threads.

The second observation is that our framework running on the UPMEM system outperforms the CPU baselines in the majority of cases, even when data transfer time is included.
The speedup achieved over the best CPU baseline is up to 4.06$\times$ in the case of SWG.
For the state-of-the-art algorithms, WFA and WFA-adaptive, the speedups achieved over the best CPU baseline are up to 1.83$\times$ and 2.56$\times$, respectively.
These results demonstrate the effectiveness of PIM at accelerating memory-bound sequence alignment workloads.

Here, we would like to reiterate that our CPU baselines are powerful server-grade dual-socket systems running at full scale.
CPU hardware and software have been developed and optimized for decades by large teams of engineers, whereas the UPMEM PIM system has been developed over a few years by a small team.
We expect that the relative advantage of PIM systems will be more pronounced as PIM hardware and software matures.
We also note that in our current system, the DPUs are running at 425~MHz, however they are expected to run at 500~MHz in future systems which would further improve performance.

The third observation is that, when the data transfer time is \textit{not} included, our framework achieves a speedup over the best CPU baseline of up to 28.14$\times$ in the case of WFA-adaptive (25.93$\times$ for WFA).
The results without transfer time are important for two reasons.
The first reason is that in the current UPMEM system, the UPMEM DIMMs cannot be used as regular main memory.
For this reason, the CPU must read the data from disk to main memory then transfer it from main memory to the UPMEM DIMMs.
However, in future systems where a PIM module could also be used as main memory, the CPU could potentially read data from disk to the PIM module directly and then perform the alignment in the PIM module without the additional transfer.
The second reason is that in the current UPMEM system, one cannot overlap writing to MRAM by the CPU and execution by the DPUs.
However, in future systems where CPU access to a PIM module may be overlapped with execution in the PIM cores, such overlapping can hide some of the latency of writing to the PIM module.
Therefore, in light of the two aforementioned reasons, the results without the transfer time demonstrate the potential that future PIM systems have for accelerating memory-bound sequence alignment workloads.

The fourth observation is that our framework does not outperform the best CPU baseline for NW and SWG at small read lengths.
NW and SWG have more regular memory access patterns than the other alignment algorithms, and when the read lengths are small, the intermediate data structures fit into the CPU cache.
The combination of these two factors makes the memory bandwidth bottleneck less pronounced, which diminishes the advantage of PIM.
However, for other algorithms where the memory access pattern is less regular, and for larger read lengths where the sizes of the intermediate data structures outgrow the size of the CPU cache, the computation becomes more memory bound causing our framework to achieve large speedups over the CPU baseline.

\begin{figure}[b]
    \centering
    \includegraphics[width=\columnwidth]{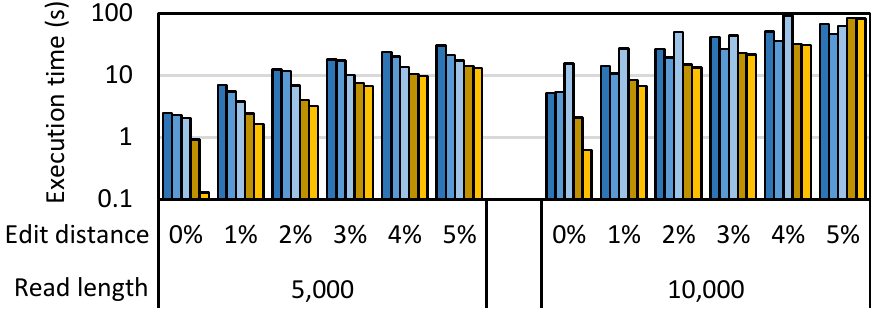}
    \caption{Execution time of our framework aligning one million sequence pairs using WFA-adaptive for large read lengths compared with CPU baselines (same legend as Fig.~\ref{fig:results-performance})}\label{fig:results-wfa-adap-large-reads}
\end{figure}

To further study the scalability of our framework, we evaluate the state-of-the-art algorithm, WFA-adaptive, on large sequence lengths of 5Kbp and 10Kbp.
NW and SWG have difficulty scaling to such large read lengths because their quadratic space complexity causes them to exceed the MRAM capacity even for a single alignment.
However, WFA and WFA-adaptive are not limited by MRAM capacity due to their linear space complexity.
The execution time of WFA-adaptive for large read lengths while aligning one million sequence pairs is shown in Fig.~\ref{fig:results-wfa-adap-large-reads}.
We observe that our framework executing on the UPMEM system continues to outperform all CPU baselines.
The only exception is for read length 10,000 with an edit distance threshold of 5\%.
In this case, the WRAM capacity is only sufficient to support a single DPU thread per DPU which underutilizes the DPU pipelines.
Although the algorithm is not limited by the 64MB MRAM for storing all the wavefront components, it is limited by the 64KB WRAM for storing the select wavefront components needed for computing a new wavefront component.
Hence, our current framework executing on the current hardware cannot scale to read lengths far beyond 10,000 for WFA and WFA-adaptive due to WRAM capacity constraints.
In some bioinformatics applications (such as minimap2), sequence alignment is typically performed only between every two seed chains to avoid long execution time and peak memory that could result from performing sequence alignment on the complete read sequence.
Thus, we believe that our framework is useful for a wide range of applications despite the sequence length limitation.
However, if sequence lengths far beyond 10,000 are a concern, the current limitation can be mitigated by streaming partial wavefront components from MRAM to WRAM in order to reduce the WRAM footprint and support aligning more sequence pairs of larger length simultaneously, or by using multiple DPU threads to align a single sequence pair.
These improvements are the subject of our future work.
We also expect the supportable read length and the performance for large read lengths to improve in future systems as the hardware and software mature, as the clock frequency of the DPUs increases, and as future systems may have larger WRAM capacity.

\subsection{PIM Performance versus GPU Performance}\label{sec:results-gpu}

Table~\ref{tab:gpu} compares the throughput of our PIM implementations of WFA-adaptive (the fastest algorithm) to a recent GPU implementation of WFA-adaptive by \cite{aguado2022wfa} using the results reported in that work.
It is clear that our PIM implementation outperforms the GPU implementation in the majority of cases.
This result shows that PIM is a promising technology for accelerating sequence alignment, even when compared to mature and widely used accelerators such as GPUs.

\begin{table}[ht]
    \centering
        \resizebox{\columnwidth}{!}{
            
\begin{tabular}{ccccc}
    \hline
    Sequence & Edit & \multicolumn{2}{c}{Throughput (alignments per second)} & Throughput \\
    \cline{3-4}
    length & distance & WFA-GPU & UPMEM (with transfer) & improvement \\
    \hline
    150 & 2\% & 9.09M & 12.97M & 1.42$\times$ \\
     & 5\% & 5.56M & 7.03M & 1.27$\times$ \\
    1,000 & 2\% & 1.43M & 1.10M & 0.77$\times$ \\
     & 5\% & 370K & 434K & 1.17$\times$ \\
    10,000 & 2\% & 25.0K & 66.9K & 2.68$\times$ \\
     & 5\% & 5.56K & 11.81K & 2.12$\times$ \\
    \hline     
\end{tabular}

        }
    \caption{Comparison with WFA-GPU~\citep{aguado2022wfa}}
    \label{tab:gpu}
    \vspace{-20pt}
\end{table}

\subsection{Using WRAM Only or WRAM and MRAM}\label{sec:results-wram-vs-mram}

\begin{figure*}
    \centering
    \includegraphics[width=\textwidth]{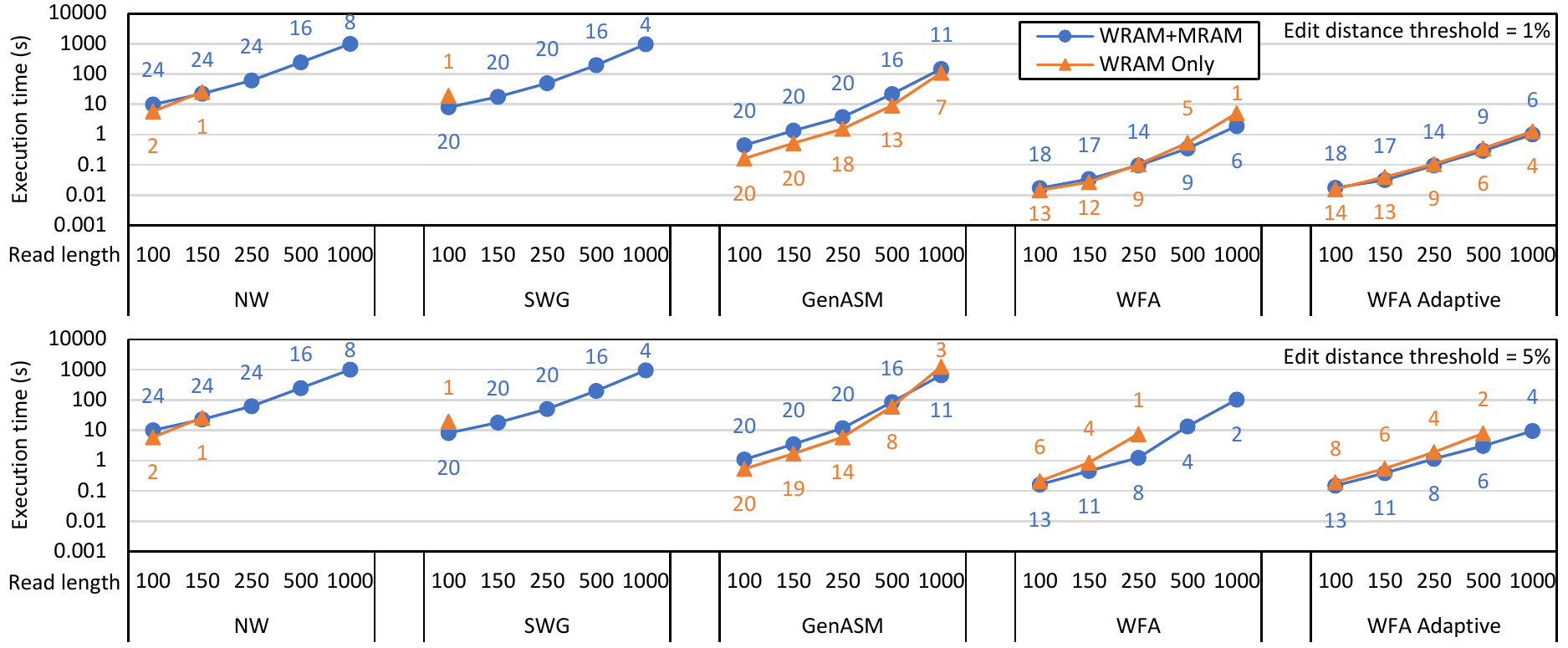}
    \caption{Execution time of our framework aligning five million sequence pairs with edit distance 1\% and 5\% using WRAM Only or WRAM and MRAM (labels indicate number of threads per DPU)
    }\label{fig:wram-vs-mram}
\end{figure*}

Recall from Section~\ref{sec:wramvsmram} that our framework provides two implementations of each algorithm: one that only uses WRAM for intermediate alignment data structures, and another that uses MRAM for these data structures and transfers data currently being accessed from these data structures to WRAM as needed.
Fig.~\ref{fig:wram-vs-mram} compares the execution time and scalability of these two implementations for each algorithm and read length with edit distance 1\% and 5\%.
Based on these results, we make three key observations.

The first observation is that for the algorithms that use a large amount of memory for the intermediate alignment data structures (i.e., NW and SWG), the implementations that use both WRAM and MRAM scale better with the read length than those that only use WRAM.
In the case of NW for small read lengths, the implementation that uses WRAM only is up to 1.70$\times$ faster.
However, for the remaining cases, this implementation is slower or cannot even execute.
The reason is that the implementations that use WRAM only can only support a small number of DPU threads due to the constrained WRAM capacity, which prevents them from utilizing the DPU pipeline well.
In contrast, the implementations that use both WRAM and MRAM can support a larger number of DPU threads, causing them to perform better for long read lengths despite incurring higher memory access latency.
Note that the trends for these two algorithms are independent of the edit distance since their operations and memory consumption are independent of the alignment score.

The second observation is that for the algorithm that uses a small amount of memory for the intermediate alignment data structures (i.e., GenASM), the implementation that only uses WRAM is usually faster (up to 2.76$\times$ for edit distance 1\%).
The reason is that the WRAM only implementation can support a large enough number of DPU threads to utilize the DPU pipeline well.
We note, however, that for larger edit distances that require more memory, the implementation that uses WRAM and MRAM becomes more favorable for large reads.
For example, for read length 1,000 and edit distance 5\%, the implementation that uses WRAM and MRAM together is 1.91$\times$ faster.

The third observation is that for the implementations that use a moderate amount of memory for the intermediate alignment data structures (i.e., WFA and WFA-adaptive), the implementations that use WRAM only are faster for shorter reads and low edit distance (up to 1.17$\times$ for WFA and 1.12$\times$ for WFA-adaptive at an edit distance of 1\%).
On the other hand, the implementations that use both WRAM and MRAM are faster for longer reads (up to 2.70$\times$ for WFA and 1.25$\times$ for WFA-adaptive at an edit distance of 1\%).
The reason is that the number of DPU threads that can be used decreases as the read length gets larger, which favors the implementation that can use more threads over the one with lower access latency.
Note that the difference between the two implementations grows as the edit distance grows because the memory consumption of WFA and WFA-adaptive is highly sensitive to the edit distance.
For example, for edit distance 5\%, the implementation that uses WRAM and MRAM together is up to 6.15$\times$ faster.

The three observations presented in this section demonstrate the importance of our framework supporting both implementations of each algorithm, those that use WRAM only and those that use WRAM and MRAM for intermediate alignment data structures.
In either case, our framework can automatically identify the maximum number of threads that can be used to alleviate this burden from the user.

\vspace{-10pt}

\section{Conclusion}\label{sec:conclusion}

We present a framework for high-throughput pairwise sequence alignment that overcomes the memory bandwidth bottleneck by using real processing-in-memory systems.
Our framework targets UPMEM, the first publicly-available general-purpose programmable PIM architecture.
It supports multiple alignment algorithms and includes two implementations of each algorithm that manage the UPMEM memory hierarchy differently and are suitable for different read lengths.
Our evaluation shows that our framework executing on an UPMEM PIM system substantially outperforms parallel CPU implementations executing at full-scale on dual-socket server-grade CPU systems.
Our results demonstrate that PIM systems provide a promising alternative for accelerating sequence alignment.
We expect even larger improvements from future incarnations of PIM systems.

\vspace{-5pt}

\section*{Acknowledgements and Funding}
We thank Fabrice Devaux and Rémy Cimadomo for their valuable support, including insightful feedback and access to UPMEM hardware.
This work has been supported by the University Research Board of the American University of Beirut [URB-AUB-104107-26306].
We also acknowledge the generous gifts provided by the SAFARI Research Group's industrial partners, including ASML, Facebook, Google, Huawei, Intel, Microsoft, VMware, and Xilinx, as well as the support from the Semiconductor Research Corporation, the ETH Future Computing Laboratory, and the European Union’s Horizon programme for research and innovation under grant agreement No 101047160, project BioPIM (Processing-in-memory architectures and programming libraries for bioinformatics algorithms).

\balance
\bibliographystyle{IEEEtran}
\bibliography{main}

\end{document}